\documentclass[useAMS,usenatbib]{mn2e}
\usepackage{graphicx}
\usepackage{amsmath}
\usepackage{amssymb}
\usepackage{bm,url,times}
\usepackage{float}
\usepackage{xcolor}

\def\Pm{\mbox{P}_M}
\def\Rm{\mbox{R}_M}

\def\Rey{\mbox{Re}}
\def\urms{{ u_{rms}}}

\def\kf{{k_f}}
\def\ted{{t_{ed}}}

\def\aap{A\&A}
\def\apjl{ApJ}

\def\apjs{ApJS}

\newcommand{\EQ}{\begin{equation}}
\newcommand{\EN}{\end{equation}}
\newcommand{\EQA}{\begin{eqnarray}}
\newcommand{\ENA}{\end{eqnarray}}

\newcommand{\Eq}[1]{Eq.~(\ref{#1})}
\newcommand{\Eqs}[2]{Eqs.~(\ref{#1}) and~(\ref{#2})}

\newcommand{\Fig}[1]{Fig.~\ref{#1}}

\newcommand{\bra}[1]{\langle #1\rangle}

{}
{}
{}

{}
{}
{}
{}
{}
{}
{}
{}
\newcommand{\meanAA}{\overline{\mbox{\boldmath $A$}}{}}{}
\newcommand{\meanBB}{\overline{\mbox{\boldmath $B$}}{}}{}
{}
{}
{}
{}
{}
{}
{}
{}
{}
\newcommand{\meanJJ}{\overline{\mbox{\boldmath $J$}}{}}{}
{}
{}

{}
{}
{}

{}

{}
{}
\newcommand{\emf}{{\cal E}}{}

\newcommand{\eee}{\hat{\mbox{\boldmath $e$}} {}}

\newcommand{\xxx}{\hat{\mbox{\boldmath $x$}} {}}

\newcommand{\zzz}{\hat{\mbox{\boldmath $z$}} {}}

\newcommand{\kk}{\bm{k}}

\newcommand{\xx}{\bm{x}}

\newcommand{\bb}{\bm{b}}
\newcommand{\BB}{\bm{B}}

\newcommand{\uu}{\bm{u}}

\newcommand{\jj}{\mbox{\boldmath $j$} {}}
\newcommand{\JJ}{\mbox{\boldmath $J$} {}}

\newcommand{\AAA}{\mbox{\boldmath $A$} {}}
\newcommand{\aaaa}{\mbox{\boldmath $a$} {}}

\newcommand{\ff}{\mbox{\boldmath $f$} {}}

\newcommand{\nab}{\mbox{\boldmath $\nabla$} {}}

\def\la{\mathrel{\mathchoice {\vcenter{\offinterlineskip\halign{\hfil
$\displaystyle##$\hfil\cr<\cr\sim\cr}}}
{\vcenter{\offinterlineskip\halign{\hfil$\textstyle##$\hfil\cr<\cr\sim\cr}}}
{\vcenter{\offinterlineskip\halign{\hfil$\scriptstyle##$\hfil\cr<\cr\sim\cr}}}
{\vcenter{\offinterlineskip\halign{\hfil$\scriptscriptstyle##$\hfil\cr<\cr\sim\cr}}}}}
\def\ga{\mathrel{\mathchoice {\vcenter{\offinterlineskip\halign{\hfil
$\displaystyle##$\hfil\cr>\cr\sim\cr}}}
{\vcenter{\offinterlineskip\halign{\hfil$\textstyle##$\hfil\cr>\cr\sim\cr}}}
{\vcenter{\offinterlineskip\halign{\hfil$\scriptstyle##$\hfil\cr>\cr\sim\cr}}}
{\vcenter{\offinterlineskip\halign{\hfil$\scriptscriptstyle##$\hfil\cr>\cr\sim\cr}}}}}
\def\Pm{\mbox{\rm Pr}_M}
\def\Rm{R_{\rm m}}

\def\Rey{\mbox{\rm Re}}

\def\kf{k_{\rm f}}

\def\urms{u_{\rm rms}}

\graphicspath{{./figs/}{./png/}}

\title[]
{Saturation of large-scale dynamo in anisotropically forced turbulence}
\author[]{Pallavi Bhat$^{1,2}$\thanks{
	pallavi.bhat@icts.res.in}\\
$^{1}$ International Centre for Theoretical Sciences, Tata Institute of Fundamental Research, Bangalore 560089, India \\
$^2$ Department of Applied Mathematics, University of Leeds, Leeds, UK, LS29JT}
\begin{document}

\pagerange{\pageref{firstpage}--\pageref{lastpage}} \pubyear{2020}

\maketitle

\label{firstpage}

\begin{abstract}
Turbulent dynamo theories have faced difficulties in obtaining evolution of large-scale magnetic fields on 
short dynamical time-scales due to the constraint imposed by magnetic helicity balance. 
This has critical implications for understanding the large-scale magnetic field evolution in astrophysical 
systems like the Sun, stars and galaxies. Direct numerical simulations (DNS) in the past with isotropically forced 
helical turbulence have shown that large-scale dynamo saturation time-scales are 
dependent on the magnetic Reynolds number ($\Rm$).
In this work, we have carried out periodic box DNS of helically forced turbulence leading to a 
large-scale dynamo with two 
kinds of forcing function, an isotropic one based on that 
used in \textsc{Pencil-Code} and an anisotropic one based on Galloway-Proctor flows.
We show that when the turbulence is forced anisotropically, 
the nonlinear (saturation) behaviour of the large-scale dynamo is only weakly dependent on $\Rm$. 
In fact the magnetic helicity evolution on small and large scales in the anisotropic case 
is distinctly different from that in the isotropic case.
This result possibly holds promise for the alleviation of important issues like catastrophic quenching. 
\end{abstract}


\section{Introduction}

Most astrophysical systems in the universe host coherent large-scale magnetic fields 
\citep{hathaway2015,beck2015,becketal2019}. The theoretical understanding of the origin and evolution 
of these large-scale fields is an outstanding problem in modern astrophysics. This affects also the solar-cycle, 
star-formation, evolution of inter-stellar matter, transport in accretion-disks, formation of jets near black-holes, 
to provide some examples of astrophysical scenarios where large-scale magnetic fields 
are thought to play an important role \citep{charbonneau2010,Haverkorn_2008,pudritzetal2012,bp1982}. 

Turbulent dynamo theory is the most popular paradigm used to understand the large-scale field 
generation and evolution in astrophysical systems \citep{ss21}. 
Idealized direct numerical simulations (DNS) in the past couple of 
decades have demonstrated that it is possible to obtain growth of spatio-temporally oranized magnetic fields 
in a turbulent fluid provided there is some ingredient to the turbulence which breaks mirror symmetry. 
However, a thorny issue has been one regarding the timescales. The observed magnetic 
fields are thought to be existing in saturated nonlinear state at the current epoch. The 
evolution timescales of solar or galactic large-scale fields are found to be incompatible with the existing 
dynamo theories and DNS results \citep{BS05}. 
The issue has been identified to be with the non-cooperation of magnetic helicity. 
Magnetic helicity, a topological quantity, is considered to be robustly conserved in the limit of 
large magnetic Reynolds numbers ($\Rm$) \citep{Berger1984,blackman2015}. Most astrophysical systems 
are high $\Rm$ systems (also high fluids Reynolds 
number $\Rey$). As a result, it has been found that it takes long resistive timescales for magnetic helicity to grow 
in the nonlinear regime. 
This affects the timescales of growth of the large-scale fields which are directly associated with the growth of the 
magnetic helicity \citep{axel2011,BSS2012}. 

Already in the 1990's, \citet{VC92,CH1996} argued for an $\Rm$ dependent saturation behaviour of the large-scale dynamo. 
Known as catastrophic quenching, it was thought that the culprit is the $\Rm$ dependent small-scale field evolution 
which supresses the large-scale field. 
However, later works brought an interpretation to catastrophic quenching 
based on the magnetic helicity evolution \citep{GD94,BY95}. 
It was works by \citet{FB2002,BB2002,Kandu2002} which generalized this interpretation transforming 
the notion of quenching being catastrophic to one that was less severe, known as "dynamical quenching".
Nonetheless, it was identified that in the $\alpha^2$ dynamo case, the steady state limit of dynamical 
quenching leads to catastrophic quenching in the turbulent transport coefficients \citep{BS05}.
Ultimately, this result is rooted in the fact that the timescales of evolution of the magnetic helicity are 
resistively constrained and thus the DNS routinely show long timescales for growth of the large-scale fields 
in the nonlinear regime. 
In this paper, this issue will be referred to simply as "$\Rm$ dependent saturation".

The suggested solution to the issue of $\Rm$ dependent saturation was to allow for magnetic helicity 
fluxes in and out of the given system 
\citep{BF2000}. 
This led to a search for helicity fluxes in the DNS to alleviate the $\Rm$ dependence 
\citep{Mitraetal2010,DelSordoetal2013,B2018}. 
This endeavour has been largely unsuccessful given that even at fairly large $\Rm$, 
the fluxes were found to be not yet sufficiently significant. However, the simulations with open boundaries 
tend to show more positive results compared to those with closed boundaries \citep{KKB2010}. 
Recent works on this \citep{CBT2020,Bhatetal2021} found 
that in their open domain, the resistive term remained the dominant one up until the saturation of magnetic helicity 
and the saturation consisted of the fluxes simply opposing the resistive term and thus, opposing the growth 
of the magnetic helicity. 

The simplest DNS giving rise to a large-scale dynamo, the $\alpha^2$-dynamo, typically employs isotropic helical forcing 
and no shear \citep{B2001,BSB2016}. 
The ones that employed instead anisotropic forcing, either included also uniform shear or had a non-conventional parameter
regime of small $\Rey$ \citep{sv2011,PNCT2016,CBT2020}. 
In this paper, we perform simulations leading to the simple $\alpha^2$ dynamo using the same setup as \citet{Bhatetal2021}, 
but the large-scale dynamo is in the more conventional regime of 
large $\Rm$ and reasonably large $\Rey$. We find that during the kinematic regime the generated turbulence is isotropic 
despite the anisotropic nature of forcing (due to a sufficiently large $\Rey$). However, when the dynamo saturates, the 
manifestation of the anisotropy of the forcing leads to a saturation behaviour that is different from the previous cases 
and potentially independent of $\Rm$ (or only weakly dependent on $\Rm$). 

The paper is organized as follows. Section~\ref{numset} specifies the numerical setup including the model 
and the two different kinds of forcing function that have been used for comparison purposes. Section~\ref{results} 
details the results from our DNS, where we first show the behaviour of total magnetic helicity for different $\Rm$ 
and then study the evolution of the total current helicity as well. The main result is in subsection~\ref{effectaniso}. 
Explanations for this result are offered in the subsection~\ref{satbeh}. Finally, we summarize the results and discuss
the relevance of our work to astrophysical settings and briefly mention future work in \ref{summa}.

\section{Numerical setup}
\label{numset}
\label{setup}
 
\subsection{\label{model}The model}
The momentum equation and the induction equation are solved on a cartesian grid. 
The fluid is incompressible and the density is taken to be unity. We solve for the vector potential 
instead of the magnetic field by employing the winding gauge \citep{PY2014}. 
The equations are given by, 
\EQ 
{\partial\uu\over \partial t}= -\uu \cdot \nab \uu - \nab P + {\JJ\times\BB}
+ {\bm F} + \nu\nab^2 \uu \, 
\label{mom} 
\EN
\EQ
{\partial \AAA\over \partial t}= \uu\times\BB - \eta \JJ + \nab \phi \, 
\label{ind}
\EN
where $\uu = (u,v,w)$ is the velocity field and $\BB=\nab \times \AAA$ is the magnetic field. 
The momentum equation (\ref{mom}) and induction equation (\ref{ind}) are subject to 
the constaints of incompressibility, $\nab \cdot \uu=0$ and winding gauge, 
$\nab_H \cdot \AAA_H=0$ respectively, where the subscript $H$ denotes the 
horizontal component i.e. along $(\xxx,\zzz)$. The current density is $\JJ=\nab \times \BB$. The viscosity and resistivity are denoted by $\nu$ and $\eta$ respectively.
${\bm F}$ is the forcing function which injects energy into the flow. 

\subsection{Forcing functions}
We have employed two types of forcing functions. 
The first is modelled using 2.5-dimensional cellular flows known as Galloway-Proctor flow
and we will refer to it as GP forcing. 
The second is one that is used routinely in \textsc{Pencil-Code}\footnote{http://pencil-code.nordita.org} and will be referred to as PC forcing. 
We describe these below. 

\subsubsection{Galloway-Proctor forcing function}

The GP forcing is given by ${\bm F}=\sum_k A_k\left(\partial_y\psi_k,-\partial_x \psi_k,\chi_k \right)$ \citep{TC2013}.
Here, the streamfunction $\psi_k$ is given by, 
\EQA
\psi_k(x,y,t)&=& \sin{k\left((x-\zeta_k)+\cos{(\omega_kt)}\right)} \\ \nonumber
&+&\cos{k\left((y-\eta_k) +\sin{(\omega_kt)}\right)}.
\ENA
The vertical velocity is $\chi_k=k \psi_k(x-\gamma_k, y-\delta_k,t)$. The offsets $\zeta_k$, $\eta_k$, $\gamma_k$ and $\delta_k$ are functions of time consisting of piece-wise constant random sequences with range $(0,2\pi)$. Note that only when $\gamma_k=0$ and $\delta_k=0$, maximally helical flows are obtained. Note that the flow is independent of the coordinate $z$.
The range of wavenumbers at which the flow is forced is $k=8$--$9$. Thus the average
forcing wavenumber is around $k_f\sim8.5$. 
The turbulent flow generated by this forcing reflects anisotropy in the root mean squared (RMS) values of 
the different components in velocity. While the RMS of $u$ and $v$ are similar, RMS of $w$ is $\sim$1.4 times 
that of $u$ or $v$. But this degree of anisotropy is not sufficiently high enough to lead to a power spectrum different from one with Kolmogorov slope of 
$k^{-5/3}$ in the inertial range. 

\subsubsection{Pencil-Code forcing function}

The PC forcing function is given by ${\bm F} = \ff(\xx,t)$, which is random in time
and defined as \citep{B2001},
\EQ
\ff(\xx,t)=\mbox{Re}\{N\ff_{\kk(t)}\exp[i\kk(t)\cdot\xx+i\phi(t)]\},
\EN
where $\kk(t)$ is a time dependent wavevector,
and $\phi(t)$ with $|\phi|<\pi$ is
a random phase. The normalization factor
$N$ is inversely proportional to ${\delta t}^{1/2}$, where $\delta t$ is the length of the
timestep.  At each timestep randomly one of the many possible vectors
in $8<|\kk|<9$ is selected. The system is forced with eigenfunctions of the
curl operator,
\EQ
\ff_{\kk}={\kk\times(\kk\times\eee)-i|\kk|(\kk\times\eee)
\over2\kk^2\sqrt{1-(\kk\cdot\eee)^2/\kk^2}},
\EN
where $\eee$ is an arbitrary unit vector needed in order
to generate a vector $\kk\times\eee$ that is perpendicular
to $\kk$.  This constitutes the maximally helical forcing.

~\\
To carry out the numerical simulations, we employ the open source code, 
\textsc{Dedalus}\footnote{https://dedalus-project.org} \citep{dedalus}.
It is a pseudo-spectral code which has a flexible framework to simulate custom equations and is MPI parallelized. 

\subsection{Simulation setup}
We have carried out all the simulations in a domain with periodic boundaries. 
Three different values of the magnetic Reynolds numbers have been explored (150, 300, 600) 
corresponding to three different resolutions of $(64\times64\times32)$,
$(128\times128\times64)$ and $(256\times256\times128)$ respectively.  The magnetic Reynolds number is defined as,
$\Rm=\urms (2\pi/k_f)/\eta$, where $k_f$ is the average forcing wavenumber.

The velocity is initially zero and the magnetic field is initialized to be random and weak. 
In all the simulations, the $\urms$ is roughly of the order of unity. 
The eddy turn over timescale is defined as $t_{ed}=2\pi/(\urms \kf)$.  

\section{Results}
\label{results}
\begin{figure}
\includegraphics[width=0.45\textwidth, height=0.22\textheight]{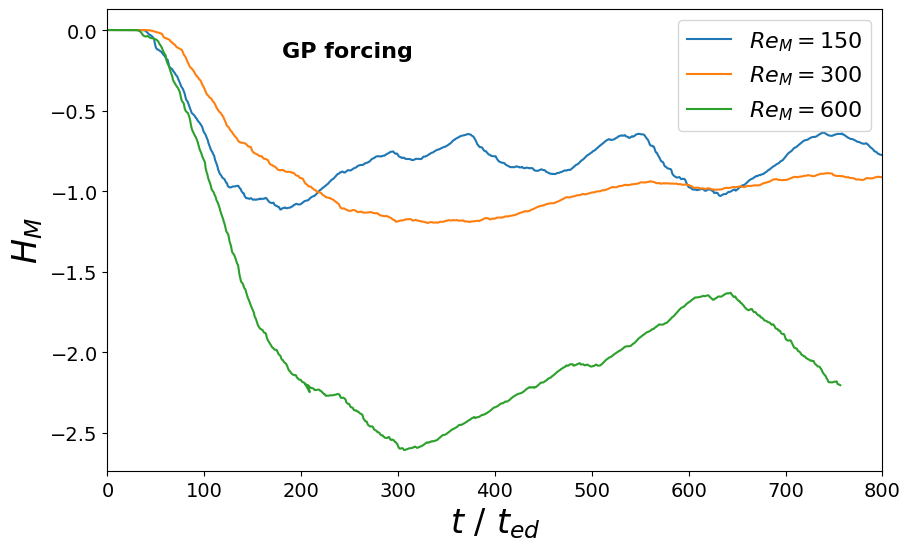}
\includegraphics[width=0.45\textwidth, height=0.22\textheight]{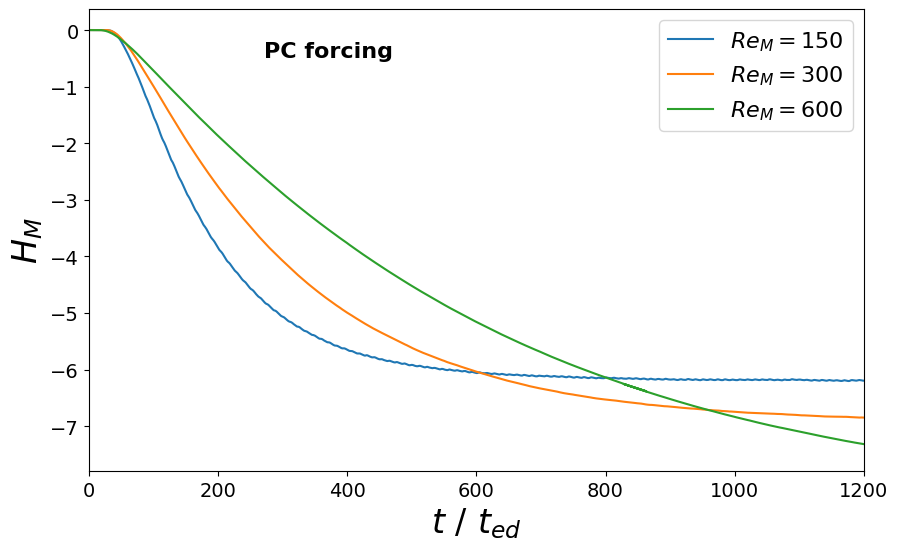}
\caption{
Evolution of magnetic helicity is shown from three different runs with $\Rm=150$, $300$ and $600$ respectively.
The upper panel is for runs with GP forcing and the lower panel is for runs with PC forcing. 
}
\label{helRm}
\end{figure}

\subsection{Magnetic helicity evolution}
\label{maghelevol}
It has been shown that in a helical turbulence simulation, the generation of 
large-scale fields is associated with the generation of magnetic helicity, $H_M=\bra{\AAA \cdot \BB}$, 
where $\bra{}$ represents the total volume average \citep{BS05}. 
Thus, the evolution of magnetic helicity is of prime importance for the large-scale dynamo. 
Here, in this subsection, we show the evolution of the total magnetic helicity from two sets of runs: 
ones with GP forcing and others with PC forcing. The runs with the same $\Rm$ also have the same $\Pm$, 
where $\Pm \geq 1$. This is because $\Rey$ is fixed at a value of $150$ in all the runs, while 
the $\Rm$ varies from $150$ to $600$.  
In \Fig{helRm}, we show the evolution of $H_M$ with time 
at different $\Rm=150, 300$ and $600$, from the runs GP150, GP300 and GP600 in the upper panel 
and from runs PC150, PC300 and PC600 in the lower panel, respectively. 
Clearly, the runs with GP forcing have a behaviour thats different from that with PC forcing. 
In both cases, $H_M$ decays or rather grows "negatively".
While runs with PC forcing show a saturation behaviour thats $\Rm$ dependent (i.e. the time taken to saturate, $t_{sat}$  
gets increasingly long with increasing $\Rm$), such behaviour in the runs with 
GP forcing is intriguingly absent or at best has a weak dependence with $\Rm$. 
In the runs with PC forcing, with increasing $\Rm$, the growth rate of $H_M$ gets smaller whereas 
in the runs with GP forcing, the growth rate remains roughly similar irrespective of the $\Rm$ thus 
leading to a lack of a strong $\Rm$ dependence in saturation.

Before we proceed to investigate the $\Rm$ dependence (or lack of it) in the GP forcing runs, 
consider the magnetic helicity equation in a closed domain (like the periodic box), 
\EQ
\frac{d \bra{\AAA\cdot\BB}}{d t} = - 2\eta \bra{\JJ\cdot\BB}.
\label{maghel}
\EN

The typical argument presented to demonstrate the conservation of magnetic helicity for very large $\Rm$ is the following. 
In a turbulent system, the magnetic field develops sufficiently large gradients such that the term responsible for 
magnetic energy decay, $-2\eta\bra{\JJ^2}$ is finite and non-zero in the limit of resistivity tending to zero (large $\Rm$). 
Thus we have that $\JJ \propto \eta^{-1/2}$ and then the term on RHS in \Eq{maghel}, $\eta \bra{\JJ\cdot\BB}$, is estimated as being proportional to 
$\eta^{1/2}$ which goes to zero in the large $\Rm$ limit. 
Thus, the time evolution of magnetic helicity is considered to 
generally happen on long resistive timescales. Since we find that this is not true for runs with GP forcing
i.e. $t_{sat}$ is much shorter than the resistive timescale, 
it is of interest to also study the evolution of current helicity, $\bra{\JJ\cdot\BB}$. 
We do this in the next subsection.  

\subsection{Current helicity evolution}
\label{curhelrole}
\begin{figure}
\includegraphics[width=0.45\textwidth, height=0.25\textheight]{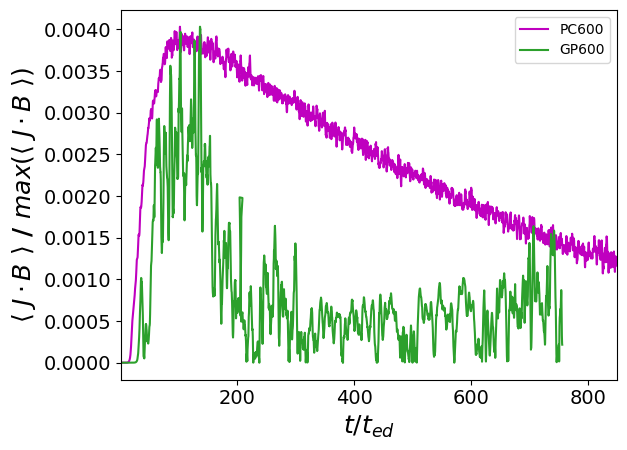}
\caption{
	The evolution of $\vert\bra{\JJ\cdot\BB}\vert$ normalized by its maximum value is shown for the runs GP600 and PC600. 
}
\label{curhelevol}
\end{figure}

\begin{figure}
\includegraphics[width=0.45\textwidth, height=0.22\textheight]{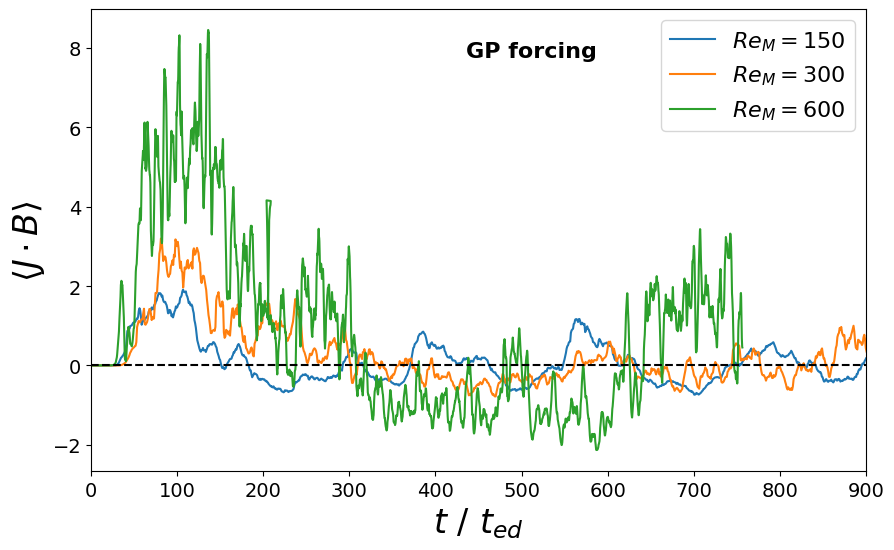}
\includegraphics[width=0.45\textwidth, height=0.22\textheight]{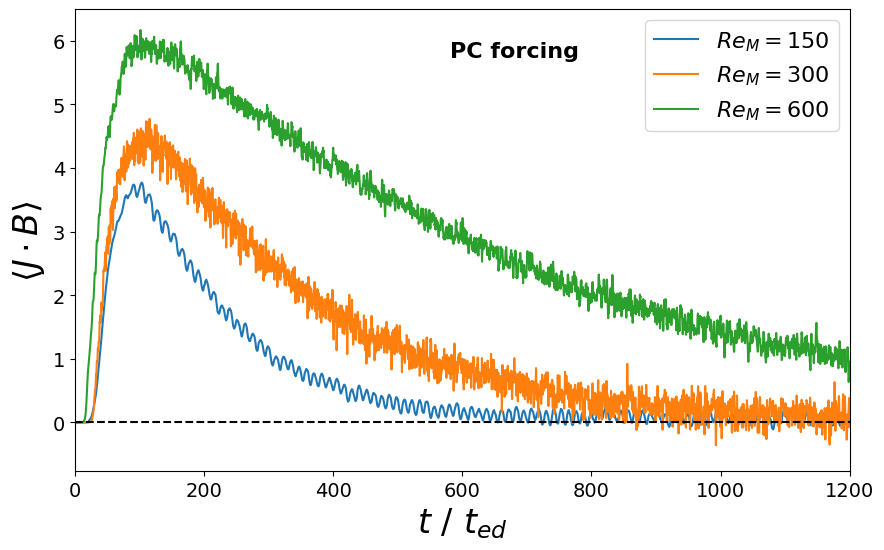}
\caption{
The evolution of the current helicity $\bra{\JJ\cdot\BB}$ is shown for three different 
	runs with $\Rm=150$, $300$ and $600$ respectively.
The upper panel is for runs with GP forcing and the lower panel is for runs with PC forcing.
}
\label{curhelevol2}
\end{figure}

In \Fig{curhelevol}, we present the evolution of the absolute value of the total 
current helicity $\bra{\JJ\cdot\BB}$ (which is not a positive definite quantity) 
for the two runs, GP600 and PC600. 
The value of $\bra{\JJ\cdot\BB}$, in both the runs, first grows and after a 
reaching a maximum, decays to zero. While the value of zero indicates saturation in magnetic helicity,  
the peak in the curve occurs some time after the saturation of the small-scale dynamo i.e. after the magnetic energy at 
large wavenumbers in the magnetic spectrum has saturated. 
It can be seen that the decay rate of $\bra{\JJ\cdot\BB}$ is much smaller in the run PC600 compared to 
that in GP600. 
Turns out that the current helicity also shows an $\Rm$ dependent behaviour in their rate of decay or decay timescales. 
In the lower panel of \Fig{curhelevol2}, 
we can see that in the case of runs with PC forcing the time taken for the current helicity to go to zero, $t_{sat}$ 
increases with $\Rm$ reflecting long resistive timescales. In the runs PC150, PC300 and PC600,  we find that 
$t_{sat}\sim500$, $800$ and $>1200$ respectively. Note that these are consistent with that estimated from 
the lower panel of \Fig{helRm}. 

Such $\Rm$ dependence is weak in the runs with GP forcing as seen in the upper panels 
of \Fig{curhelevol2}. In the runs GP150, GP300 and GP600,  we find that 
$t_{sat}\sim150$, $200$ and $250$ respectively.
It is this $\Rm$ dependence  
that is reflected in the evolution of $H_M$. While from the argument provided above (for magnetic helicity conservation), 
the strong $\Rm$ dependence is expected, the absence of it in the runs with GP forcing is surprising. 

\begin{figure}
\includegraphics[width=0.45\textwidth, height=0.25\textheight]{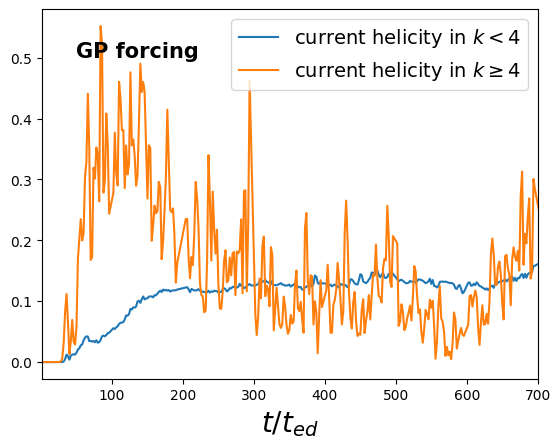}
\includegraphics[width=0.45\textwidth, height=0.25\textheight]{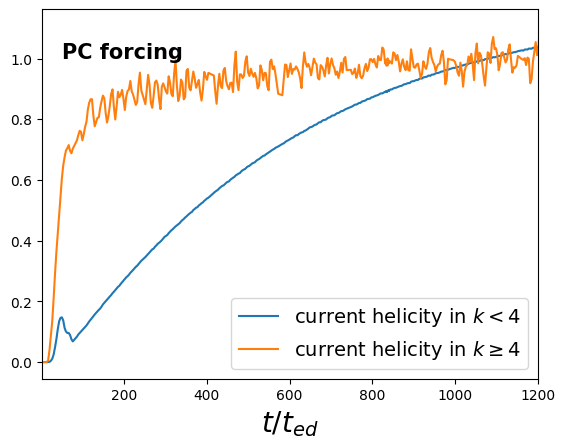}
\caption{
	The evolution of current helicity in wavenumbers $k \geq 4$ (small-scale) and $k < 4$ (large-scale) 
	is shown for the runs, GP600 and PC600 in upper and lower
panels respectively.
}
\label{curhelsplit}
\end{figure}

To understand better the evolution of the total current helicity, we split it along the 
power spectrum to assess the current helicity in large-scales versus that in the small scales. 
In the \Fig{curhelsplit}, we provide the evolution 
of current helicity in wavenumbers $k \geq 4$ and $k < 4$ for the same runs (obtained by integrating 
the current helicity power spectrum over the specified ranges of wavenumbers), GP600 and PC600 in upper and lower
panels respectively. 
In PC600 run, after small-scale dynamo saturation at around $t/\ted\sim30$--$40$ (not shown), 
the small-scale current helicity builds up 
and goes into quasi-steady state by around $t/\ted=100$ 
while the large-scale current helicity continues to grow. The point at which the two curves meet is 
when the total current helicity goes to zero. Note that we 
are plotting the absolute value of the current helicity.
And given that current helicity is not positive-definite, the sign of the small-scale current helicity is 
opposite to that of the large-scale current helicity. It can be seen that the 
time of intersection of the two curves approximately matches with 
the time at which the total current helicity goes to zero in \Fig{curhelevol}. 
Thus, the split at $k=4$ is justified to differentiate 
between large and small-scales. 

Next consider the current helicity evolution on large and small scales 
in the run GP600 as shown in the upper panel of \Fig{curhelsplit}. Here, the evolution is different from the PC600 case. 
After small-scale dynamo saturation at around $t/\ted\sim 35$--$40$ (not shown), the small-scale current helicity 
first builds up to peak around $t/\ted\sim 100$
and then starts decaying instead of acquiring steady state, thus, allowing for this curve to 
intersect with the curve for large-scale current helicity much earlier in time than in the PC600 case. 
This is what leads to the surprising fast saturation of the magnetic helicity. 

At this point, it is still not clear what is responsible for the different behaviour in the two runs with the 
same parameters of $\Rm$, $\Rey$ and $\kf$. To throw light on this matter, we next take up the issue of the 
difference in the nature of forcing function. 

\subsection{Effect of anisotropy of the forcing function}
\label{effectaniso}
\begin{figure*}
\includegraphics[width=0.87\textwidth, height=0.28\textheight]{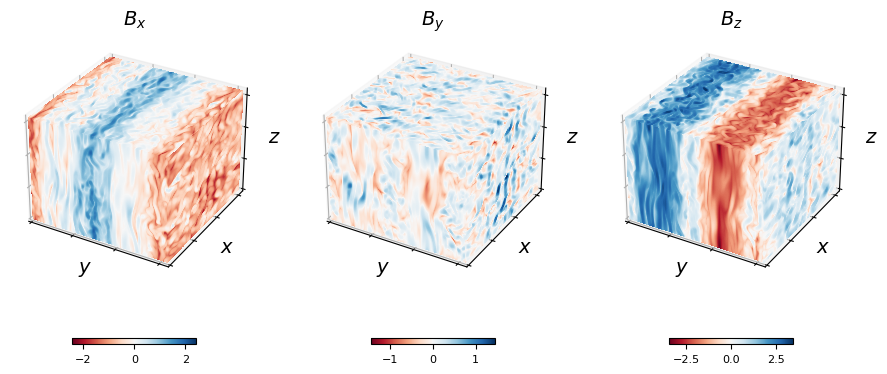}
\includegraphics[width=0.87\textwidth, height=0.28\textheight]{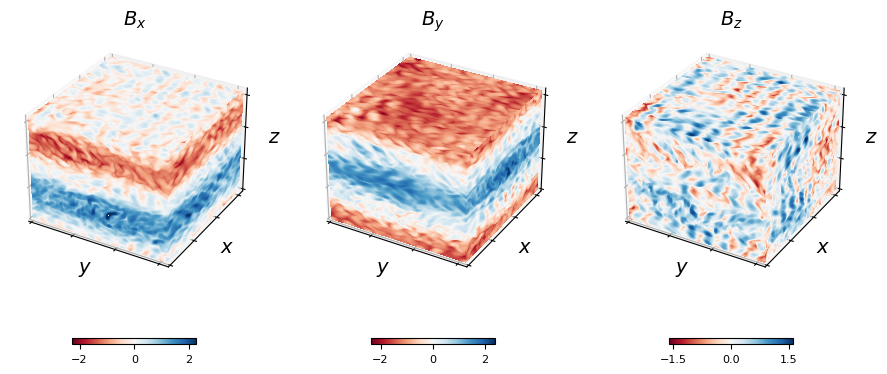}
\caption{
	The 3D field in $B_x$, $B_y$ and $B_z$ at around 
	$t/\ted=340$ is shown for the runs GP600Pm1a and GP600Pm1b in the upper and lower panels respectively.
	The run GP600Pm1a has a large-scale field in the $x$-$z$ plane wheareas in the run GP600Pm1b, the large-scale 
	field comes up in the $x$-$y$ plane. 
}
\label{boxes3d}
\end{figure*}

\begin{figure}
\includegraphics[width=0.45\textwidth, height=0.25\textheight]{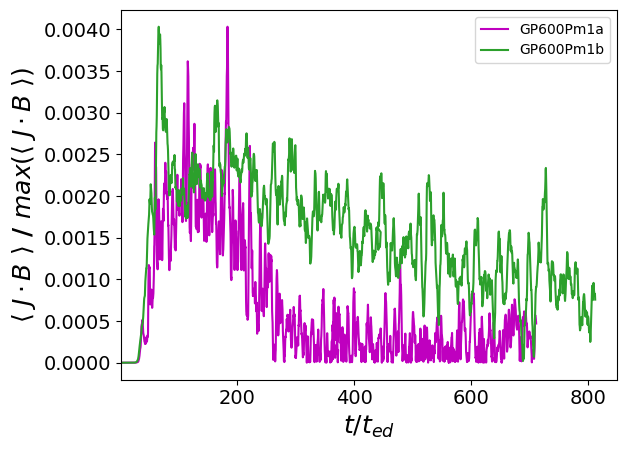}
\caption{
	The evolution of $\vert\bra{\JJ\cdot\BB}\vert$ normalized by 
	its maximum value is shown for the runs GP600Pm1a and GP600Pm1b.
}
\label{compcur3}
\end{figure}

As mentioned earlier, while PC forcing is isotropic in nature, the GP forcing is anisotropic, i.e. the 
forcing is a function of only $x$ and $y$ and not $z$. Thus the generated vortices mostly lie on the $x$-$y$ plane with the 
the helical action in the $z$-direction.
However, the anisotropy of GP forcing doesn't seem to matter 
in the kinematic phase because the large-scale field (a Beltrami mode) that arises in this system   
is one that spontaneously breaks symmetry to choose only one of the three coordinates axes to vary along or equivalently, 
one of the three planes to lie in (similar to what occurs in the runs with isotropic forcing). 
This is possible only if the underlying turbulence displays no preference for any direction. Thus, we can conclude that 
in the kinematic regime, even though the forcing is intrinsically anisotropic, the turbulence generated is effectively 
isotropic. 

Runs with both types of forcing give rise to large-scale fields that are Beltrami modes residing in the wavenumber $k=1$ 
\citep{B2001,Bhatetal2021}. 
And as mentioned before, such a large-scale Beltrami mode varies only in one of the three coordinate axes, 
which is chosen randomly. 
In the \Fig{boxes3d}, we show results from two simulations, GP600Pm1a and GP600Pm1b that are identical 
in terms of setup and parameters. 
However, one of them, run GP600Pm1b, has the large-scale 
Beltrami mode varying along $z$ (and lies in the $x$-$y$ plane) as seen in the lower panels of \Fig{boxes3d} 
and the other one, run GP600Pm1a, has the Beltrami mode varying along $y$ (and lies in the $x$-$z$ plane) 
(in the upper panels of \Fig{boxes3d}). 

In \Fig{compcur3}, we show the corresponding evolution of the current helicity in these two runs. 
The current helicity goes to zero 
in the run GP600Pm1a faster than it does in the run GP600Pm1b! This is intriguing because both the runs have identical 
setup but 
produce different results. The reason to this difference lies in the fact that the large-scale field that comes up in the 
two runs are oriented differently w.r.t the direction of the forcing function which seems to affect the dynamics
in the nonlinear regime. Thus the anisotropy of the forcing 
function seems to affect the manner of saturation. In fact the saturation behaviour in GP600Pm1b is similar to that of PC600, 
where the saturation happens on long timescales. And thus if the large-scale field is in $x$-$y$ plane, the saturation 
timescale is strongly $\Rm$ dependent as in the isotropic forcing case. We have seen this 
behaviour ($t_{sat}$ dependent on the orientation of the large-scale field that arises spontaneously) 
emerge across all our runs with GP forcing systematically. 
We detail this behaviour further in the next subsection and provide an explanation for the same. 

\subsection{Understanding the dynamo saturation behaviour} 
\label{satbeh}
We would like to understand how the difference in the saturation behaviour depends on the orientation 
of the large-scale field when the forcing is anisotropic. 
To do so, consider the equations for magnetic helicity 
for large-scale and small-scales (which can be obtained by splitting the variables like magnetic field and 
vector potential into mean and fluctuating quantities and application of Reynolds rules \citep{BS05}), 

\EQA
\label{lsh}
\frac{d \bra{\meanAA\cdot\meanBB}}{d t} &=& 2\bra{\emf\cdot\meanBB} - 2\eta \bra{\meanJJ\cdot\meanBB} \\ 
\label{ssh}
\frac{d \bra{\aaaa\cdot\bb}}{d t} &=& -2\bra{\emf\cdot\meanBB} - 2\eta \bra{\jj\cdot\bb} 
\ENA
where the overbar on the variables refers to mean quantities and the small letters 
refer to the fluctuating quantities. Also, $\emf = \overline{\uu \times \bb}$ is the electromotive force 
which is responsible for the large-scale (or mean field) growth. 

In our simulations, given the kinetic helicity generated by the turbulent forcing is of positive sign, 
the corresponding magnetic counterpart on the small-scales turns out to be positive too. And as a result, 
the sign of helicity (both magnetic and current) 
on the large-scales is negative (a direct consequence of the conservation of magnetic helicity). 
As seen in \Fig{helRm} earlier, the total magnetic helicity decays or grows "negatively" (the negative helicity 
is in excess of positive helicity). This implies 
that the total current helicity must be positive (from the \Eq{maghel}). 
This is consistent with the expectation that the total current 
helicity emphasizes the amplitude in small scales since for any given scale, 
current helicity is $k^2$ times larger than the corresponding magnetic helicity. 

However, when we split the magnetic helicity evolution into large and small scales as in \Eqs{lsh}{ssh}, the 
resistive terms are now responsible for destruction and not growth as opposed to \Eq{maghel}, where the total 
helicity grows in response to the resistive term. The growth of large or small-scale 
helicity is provided by the term $\bra{\emf\cdot\meanBB}$, which appears with the opposing signs in the two equations. 
This is consistent with the physical interpretation that when you writhe the field, corresponding twists 
appear on small-scales. 

Now, we will consider the two cases one by one. 

\subsubsection{Large-scale field in the $x$-$y$ plane}
This is the standard case which leads to $\Rm$ dependence in saturation timescale, $t_{sat}$. 
The explanations below are also applicable to isotropic PC forcing as in the run PC600.

\begin{figure}
\includegraphics[width=0.45\textwidth, height=0.25\textheight]{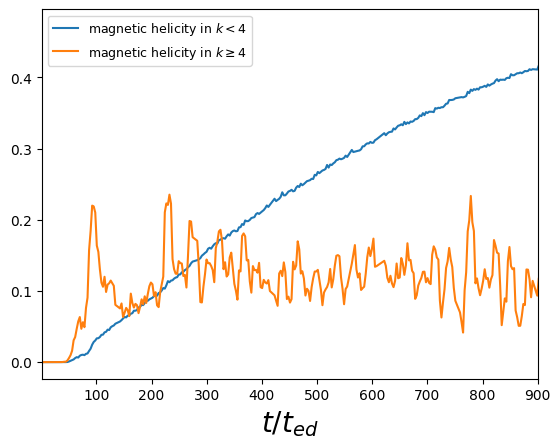}
\caption{
The evolution of magnetic helicity in wavenumbers $k \geq 4$ (small-scale) and $k < 4$ (large-scale) 
	is shown for the run GP600Pm1b.
}
\label{hel256g14}
\end{figure}

\begin{figure}
\includegraphics[width=0.425\textwidth, height=0.25\textheight]{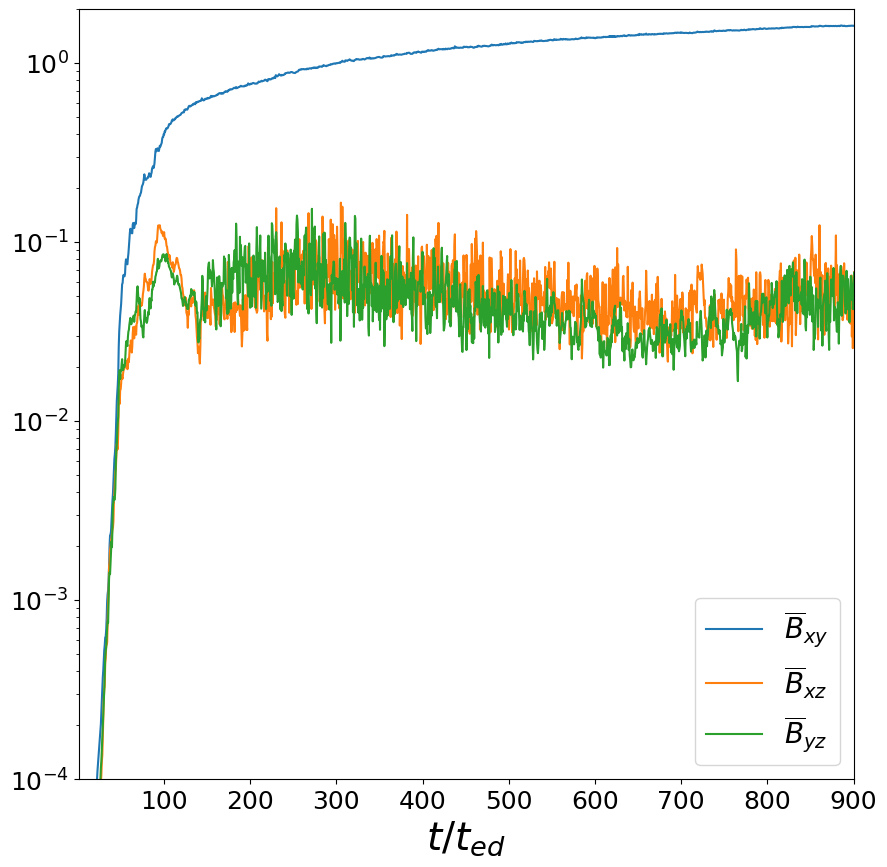}
\caption{
The evolution of large-scale fields derived from averaging in the three different planes 
is shown for the run GP600Pm1b.
}
\label{lsf256g14}
\end{figure}

In the \Fig{hel256g14}, we show the evolution of the large-scale and small-scale magnetic helicity from the run GP600Pm1b.
We find that while the small-scale helicity (SSH) has grown quickly and nearly saturated around $t/\ted=100$, 
the large-scale helicity keeps growing slowly. The timing of saturation of the small-scale helicity has certain significance as 
we explain next.
After the saturation of the small-scale dynamo around $t/\ted \sim 40$, 
the large-scale field continues to grow in, what we believe is, the quasi-kinematic 
large-scale dynamo (QKLSD) phase \citep{Bhatetal2019}. This phase shows a slope thats different from that 
in the kinematic regime as can be seen from the curve for $\meanBB_{xy}(t)$ 
(here ${\meanBB}_{xy}(t) = (1/V)\left(\int(\int \meanBB(t)~dx~dy)^2~dz\right)^{1/2}$, where $V$ is the total volume) in \Fig{lsf256g14} between 
$t/\ted \sim 40$ and $t/\ted \sim 100$. 
We believe that the growth spurt in SSH occurs during the QKLSD phase and the subsequent saturation 
of SSH marks the end of the QKLSD phase. 
The saturation of SSH implies $-\bra{\emf\cdot\meanBB} \sim \eta\bra{\jj\cdot\bb}$, 
and thus the equation for large-scale helicity can be written as 
$ d \bra{\meanAA\cdot\meanBB}/d t = -2\eta (\bra{\jj\cdot\bb} + \bra{\meanJJ\cdot\meanBB})$. 
And if we consider $\bra{\meanJJ\cdot\meanBB}\sim k_1^2 \bra{\meanAA\cdot\meanBB}$ (a fair assumption for the isotropic case), 
since $\bra{\jj\cdot\bb}$ is nearly a constant, large scale helicity (LSH) will grow on the resistive time scale (which is what we see in \Fig{hel256g14} and this has catastrophic implications at very large $\Rm$).
Eventually LSH also saturates when the growing $\bra{\meanJJ\cdot\meanBB}$ reaches the same amplitude as  
$\bra{\jj\cdot\bb}$ (only in absolute value) and they cancel out each other given their opposite signs. 

In \Fig{lsf256g14}, we find that the large-scale component in the $x$-$y$ plane, 
$\meanBB_{xy}$, grows while the other two components 
slow down at around $t/\ted=50$ and saturate by $t/\ted\sim100$. Note that the spontaneous arising of $\meanBB_{xy}$ 
already happens close to saturation of the SSD. 

The explanation above for helicity behaviour and the resistive timescales is also applicable 
to the case of isotropic forcing (and originally seen in this case) as in the run PC600. 

\subsubsection{Large-scale field in a plane perpendicular to the $x$-$y$ plane}
This is the case where the saturation time-scales are much shorter and do not exhibit a strong $\Rm$ dependence. 

\begin{figure}
\includegraphics[width=0.45\textwidth, height=0.25\textheight]{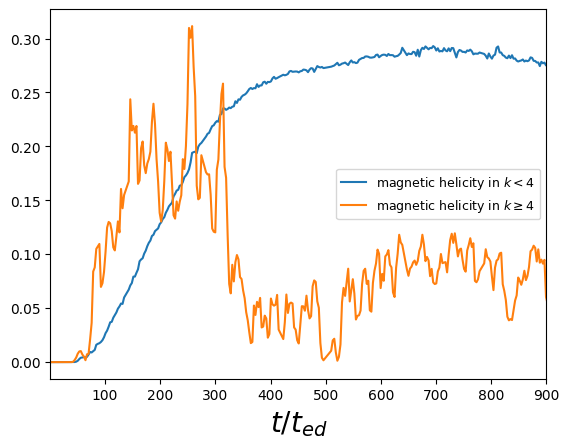}
\caption{
The evolution of magnetic helicity in wavenumbers $k \geq 4$ (small-scale) and $k < 4$ (large-scale) 
	is shown for the run GP600Pm1a.
}
\label{hel256g13}
\end{figure}

\begin{figure}
\includegraphics[width=0.425\textwidth, height=0.25\textheight]{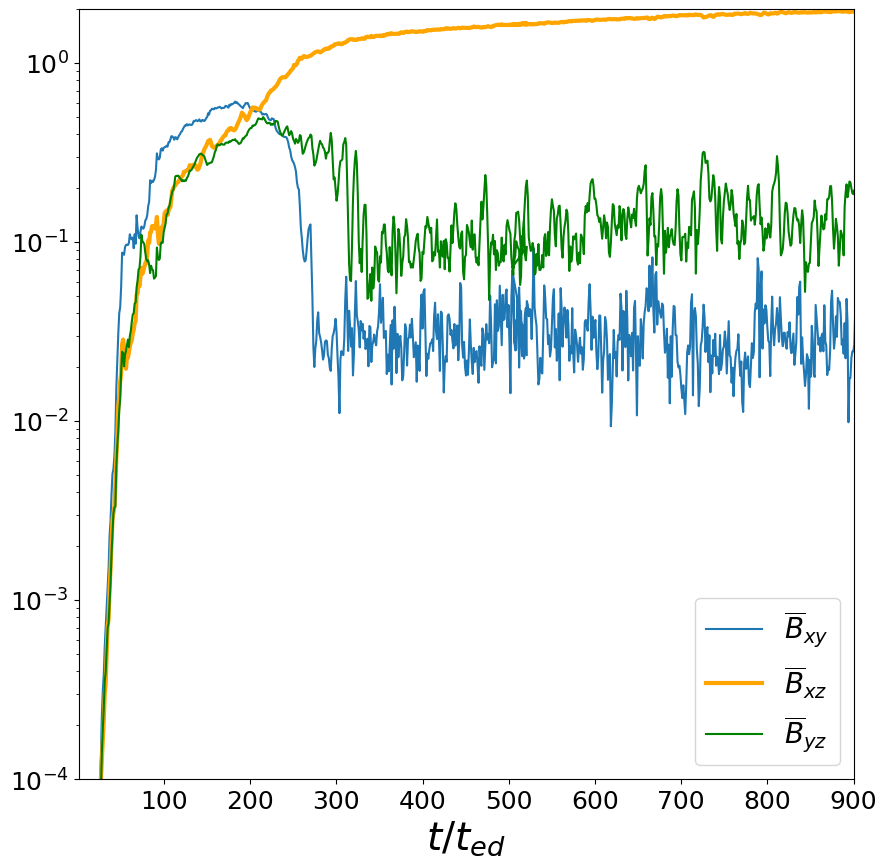}
\caption{
The evolution of large-scale fields derived from averaging in the three different planes 
is shown for the run GP600Pm1a.
}
\label{lsf256g13}
\end{figure}

In the \Fig{hel256g13}, we show the evolution of the large-scale and small-scale magnetic helicity from the run GP600Pm1a. 
We find that, in this case, the SSH builds up similar to the run GP600Pm1b but then it actually decays after growing to 
the maximum, followed by saturation (at $t/\ted\sim 320$).
On the other hand, the large-scale helicity (LSH) continues to grow as SSH decays 
and eventually saturates too (at around the same time as SSH). 
Again here the initial growth in SSH corresponds to the growth of the large-scale field in the QKLSD phase. 
This can be seen from \Fig{lsf256g13}, where the curve for ${\meanBB}_{xz}(t)$ 
(here ${\meanBB}_{xz}(t) = (1/V)\left(\int(\int \meanBB(t)~dx~dz)^2~dy\right)^{1/2}$ grows with a different slope after saturation of 
small-scale dynamo around $t/\ted \sim 50$
till nearly $t/\ted\sim 250$. This time of $t/\ted\sim 250$ corresponds to the peak/maximum in the evolution 
curve of SSH in \Fig{hel256g13}. 
The QKLSD phase becomes important in this context because the growth rate of LSH (and the corresponding growth in SSH) 
is larger in GP600Pm1a compared to GP600Pm1b! The reason for larger growth in GP600Pm1a is that the QKLSD phase 
here is much longer than that in GP600Pm1b.  This is where the alleviation of strong $\Rm$ dependence comes into play. 

Also after the QKLSD phase, the behaviour of SSH in GP600Pm1a is completely different from that in GP600Pm1b.
The subsequent decay of SSH after the maximum implies $-\bra{\emf\cdot\meanBB} < \eta \bra{\jj\cdot\bb}$ 
and the growth of LSH implies  $\bra{\emf\cdot\meanBB} > \eta \bra{\meanJJ\cdot\meanBB}$. 
Since, $\vert\bra{\jj\cdot\bb}\vert$ is larger than $\vert\bra{\meanJJ\cdot\meanBB}\vert$ before either saturates, 
the value of the source term $\vert\emf\cdot\meanBB\vert$ must be insufficient such that it decays SSH instead 
of saturating it (as in GP600Pm1b). So why is the source term smaller than that in GP600Pm1b? 

To understand this, consider the most important term that is responsible for the dynamo action from 
the induction equation, $\BB \cdot \nab \uu$. If the coherent fields generated on both large and small scales lie on the 
$x$-$y$ plane, then on expanding $\BB \cdot \nab \uu$, 
we obtain two terms. But we obtain only one term when the fields lie on the plane 
perpendicular to the $x$-$y$ plane. Considering that the derivates w.r.t $z$ are zero in the latter case, 
we obtain a reduction in the strength of the dynamo action. This would correspond to a physical picture 
where the helical action of the the flow generated by the GP forcing can more easily 
writhe or twist fields in $x$-$y$ plane as opposed to those in a plane perpendicular to the $x$-$y$ plane. 
Ofcourse, we find that this reduced dynamo action is sufficient to keep both the field and helicity growing at large-scales 
but it is insufficient to maintain those on small-scales leading to their decay.

In the \Fig{lsf256g13}, it can be seen that after the SSD saturates at around $t/\ted=40$, 
the large-scale fields in all the three 
planes grow together until around $t/\ted\sim 200$, when the ${\meanBB}_{xz}$ takes off to grow further 
and the large-scale components 
in the other two planes decay and saturate at a smaller value. 
Here the symmetry breaking happens much later as compared to that in GP600Pm1b.
The time at which the SSH starts decaying occurs after the time at 
which the large-scale component ${\meanBB}_{xz}$ arises as the dominant component (which does not coincide with 
the small-scale dynamo saturation as in GP600Pm1b). 
This supports our reasoning that the helical dynamo action is no longer as effective 
once the large-scale field, the $k=1$ Beltrami mode (and likely also a complementary small scale Beltrami mode) 
spontaneously chooses its orientation to be in a plane not as amenable to the flows 
produced by GP forcing. 

This turns out to be a blessing in disguise because as the SSH decays, the total helicity grows more negatively. 
But note that already by the time QKLSD phase ends, the growth of both the large-scale field and the large-scale helicity 
is larger than that seen in GP600Pm1b. Ultimately, both LSH and SSH saturate simultaneously because the 
decaying $\bra{\jj\cdot\bb}$ intersects with the growing curve of $\bra{\meanJJ\cdot\meanBB}$ and they cancel out 
to give $d \bra{\AAA \cdot \BB}/d t =0$. 

Now, the saturation period/timescale depends on how fast SSH (and thus small-scale current helicity) decays 
and also how fast the LSH grows. 
We find that the decay rate 
is not very different from the inital growth rate i.e. SSH decays almost as fast as it grew. 
And if the growth of SSH is independent of $\Rm$, so should its decay be. 
However, we find that the decay/growth rate has a weak dependence on $\Rm$.

\subsection{Interpretation of the results under the $\alpha$-effect formalism}

At this point, we would like to discuss the mean-field theory interpretation of $\Rm$ dependent saturation 
of the large-scale dynamo, also known as catastrophic quenching. The dynamical quenching theory incorporates 
the equation of magnetic helicity evolution to derive the net $\alpha$-effect in the nonlinear regime. 
The backreaction arises with build-up of small-scale helical fields leading to a finite 
$\bra{\jj\cdot\bb}$ term in the renormalized $\alpha$-effect formula. 
The theory then predicts slow growth of the large-scale field because 
it was thought that the only way to lose the build up 
of the small-scale helical fields (and thus decrease of $\bra{\jj\cdot\bb}$) is via their resistive decay.  

Now, if we consider the effect of anisotropic forcing on saturation as in the simulation runs of GP600 and GP600Pm1a, 
it involves fast decay of the small-scale helical fields soon after their build-up in QKLSD phase. 
And thus this would seemingly resolve the issue of $\alpha$-quenching in the standard mean field picture of 
large-scale field evolution. 

However, as we have mentioned previously, the large-scale fields arise already after 
small-scale dynamo saturation, during the QKLSD phase. Thus we find that the large-scale field strength  
is dependent on the  build-up of the helicities during the QKLSD phase. The subsequent decay of small-scale helicity 
only goes to grow the total helicity and it is not clear it affects the growth of the large-scale field itself. 
Even in the standard picture described above, it is not clear that the term $\bra{\jj\cdot\bb}$ suffers 
from a slow resistive decay (if we were to assume $\bra{\jj\cdot\bb} \sim k_f^2 \bra{\aaaa\cdot\bb}$) 
which then can aid in increasing the EMF, $\emf$. 
Instead, we find that the resistive slow growth of the large-scale helicity 
(and thus the large-scale field) is explained by 
considering $\bra{\jj\cdot\bb}$ to be constant in the equation for total helicity split into large and small scales, 
and further considering the time derivative of the small-scale helicity to be zero. 

The main difference between the saturation in isotropically forced and anisotropically forced system in the highest 
resolution runs is that the QKLSD phase lasts longer in the latter case leading to larger growth of the large-scale field. 
Currently, we do not have a clear understanding of this. In the anisotropic case, we cannot consider the simple 
form of $\alpha$-effect involving a pseudo-scalar, but instead we need to consider the full tensorial formulation \citep{rk2000}. 
Recently, \citet{HB2019} have derived the mean-field equations for the non-linear regime (including quenching) for 
exactly the case of anisotropic flows. They show how once we consider the full $\alpha$-effect (without the reduction due to 
isotropy), magnetic helicity evolution equation becomes insufficient to capture the effects of Lorentz force. Their 
results show that the quenching is worse for the anisotropic case and thus, large-scale field actually saturates at 
smaller amplitude compared to the isotropic case. This is at odds with our simulation results. 
Our comment on this is that the existing mean-field models do not capture the QKLSD phase correctly. 
In the QKLSD phase, the small-scale field has saturated and the large-scale field continues to grow (at a rate different from 
that in the SSD-dominated kinematic phase) from an amplitude that's not small. Only when the QKLSD 
phase is modelled properly, can we obtain an estimate for the time-duration of this phase and thus 
understand the difference between the cases of isotropic and anisotropic forcing. 

%
%

\section{Summary and discussions}
\label{summa}
We have carried out simulations of helically forced turbulence leading to the large-scale dynamo 
with two different forcing functions, 
an isotropic one, PC forcing and an anisotropic one, GP forcing. In the runs with PC forcing, we find the standard results 
of $\Rm$-dependent saturation (i.e. time taken to reach saturation $t_{sat}$ are long resistive timescales), 
whereas in the latter runs with GP forcing, the dependence of saturation time-scales on $\Rm$ is weak.
Thus, we find that the anisotropy of the forcing is an important factor affecting the nonlinear regime (saturation) of the 
large-scale dynamo (though it doesn't seem to be of influence in the kinematic phase, 
where the system is effectively isotropic). 
Further, even in the runs with anisotropic forcing ($z$-independent), 
when we obtain a large-scale field (a $k=1$ Beltrami mode) 
lying in the $x$-$y$ plane (${\meanBB}_{xy}$), the saturation behaviour is similar to the isotropic case. 
However, when the large-scale field 
instead lies in $x$-$z$ plane, ${\meanBB}_{xz}$ or $y$-$z$ plane, ${\meanBB}_{yz}$, 
the saturation behaviour is a weakly $\Rm$ dependent one. 

We find that the difference in the saturation behaviour arises due to the difference in the duration of the quasi-kinematic 
large-scale dynamo (QKLSD) phase. 
Runs with ${\meanBB}_{xy}$, have a short QKLSD phase followed by resistive growth of the large-scale field, taking a long 
time to saturate.
Runs with  ${\meanBB}_{xz}$ or ${\meanBB}_{yz}$ have a longer QKLSD phase leading to a larger growth in both the 
large-scale helicity and large-scale field followed by immediate saturation. Thus, we obtain shorter saturation time-scales. 
The weak $\Rm$ dependence seems to arise from this QKLSD phase 
(which has been shown to be the case in an earlier paper by \citet{Bhatetal2019}). 

The dynamics of saturation leave their signatures in the magnetic helicity evolution.
On investigating the evolution of helicities on the two different scales : small and large, we find that in both the runs, 
SSH saturates as the QKLSD phase ends. However, while in the run with ${\meanBB}_{xy}$, the SSH grows and saturates, 
in the run with ${\meanBB}_{xz}$ or ${\meanBB}_{yz}$, the SSH grows to a maximum and then decays before it saturates. 
These differences can be understood as the effect of the anisotropic forcing on the efficiency of dynamo action 
of writhing/twisting the fields based on their orientation 
and we have provided explanations for the same. 
In summary, we find that the anisotropically forced large-scale dynamo leads to a saturation behaviour 
that holds promise for alleviation of catastrophic quenching. 

We would like to consider the nature of forcing in astrophysical objects that show presence of large-scale fields. 
To grow a large-scale field, the minimum ingredient in underlying turbulence is kinetic helicity (something that breaks 
mirror-symmetry of the underlying turbulence). 
The processes which generate turbulent kinetic helicity in astrophysical objects are a combination of 
rotation and stratification. In the Sun, though the density gradient varies radially, the kinetic helicity would 
be maximum towards the poles as expected from 
the dot product of the stratification and the angular velocity \citep{Ruediger:2008}.  
In galaxies, also, the density stratification is mostly perpedicular to the rotating disk. 
In accretion disks, where the dynamos is most likely driven by magneto-rotational instability generated turbulence, 
the dynamo action is manifestly anisotropic in nature. 
Thus anisotropy of the forcing or that of the "$\alpha$-effect" can be taken as a natural ingredient to be utilized (either in 
simulations or theoretical models) to understand the behaviour of large-scale dynamos. 

An important future work would be to formulate the mean field dynamo theory for anisotropic flows, particularly  
with the consideration of QKLSD phase transitioning to the nonlinear regime and saturation. 
Further numerical studies at higher $\Rm$ are desirable. 

\section*{Data availability}
The data underlying this article were accessed from ARC4 HPC at the 
University of Leeds\footnote{https://arcdocs.leeds.ac.uk/systems/arc4.html} and 
computing clusters at International Centre for Theoretical Sciences\footnote{https://it.icts.res.in}.
The derived data generated in this research will be shared on reasonable request to the corresponding author.

\section*{Acknowledgments}
I would like to thank Prof. K. Subramanian for his very helpful discussions and comments on the paper. 
And I would like to thank Prof. S. M. Tobias for support and encouragement. 
I would like to acknowledge support by funding from the European Research Council (ERC) 
under the EU’s Horizon 2020 research and innovation programme (grant agreement D5S-DLV-786780).

\appendix

\bibliographystyle{mn2e}
\bibliography{anisoforc}

\label{lastpage}

\end{document}